\documentclass[conference,10pt]{IEEEtran}
\usepackage[english]{babel}
\usepackage{cite,authblk}
\usepackage{amsmath}
\usepackage{amssymb}
\usepackage{amsfonts}
\usepackage{graphicx}
\usepackage[colorlinks=true, allcolors=blue]{hyperref}
\usepackage{epsfig,epsf,rotating,setspace,latexsym,amsmath,amssymb,amsfonts,bm,subfigure,epstopdf}

\usepackage{amsthm}
\newtheorem{thm}{Theorem}

\theoremstyle{plain}
\newtheorem{proposition}[thm]{Proposition}

\theoremstyle{definition}
\newtheorem{example}{Example}

\newcommand{\E} {{\mathbb E}}
\renewcommand{\L} {{\mathcal L}}

\title{Age of Gossip in Networks with \\ Multiple Views of a Source}

\author[1]{Kian J. Khojastepour}
\author[2]{Matin Mortaheb}
\author[2]{Sennur Ulukus}

\affil[1]{\normalsize NYU, New York, NY}
\affil[2]{\normalsize University of Maryland, College Park, MD}

\begin{document}
\maketitle

\begin{abstract}
We consider the version age of information (AoI) in a network where a subset of nodes act as sensing nodes, sampling a source that in general can follow a continuous distribution. Any sample of the source constitutes a new version of the information and the version age of the information is defined with respect to the most recent version of the information available for the whole network. We derive a recursive expression for the average version AoI between different subsets of the nodes which can be used to evaluate the average version AoI for any subset of the nodes including any single node. We derive asymptotic behavior of the average AoI on any single node of the network for various topologies including line, ring, and fully connected networks. The prior art result on version age of a network by Yates [ISIT'21] can be interpreted as in our derivation as a network with a single view of the source, e.g., through a Poisson process with rate $\lambda_{00}$. Our result indicates that there is no loss in the average version AoI performance by replacing a single view of the source with distributed sensing across multiple nodes by splitting the same rate $\lambda_{00}$. Particularly, we show that asymptotically, the average AoI scales with $O(\log(n))$ and $O(\sqrt{n})$ for fully connected and ring networks, respectively. More interestingly, we show that for the ring network the same $O(\sqrt{n})$ asymptotical performance on average AoI is still achieved with distributed sensing if the number of sensing nodes only scales with $O(\sqrt{n})$ instead of prior known result which requires $O(n)$. Our results indicate that the sensing nodes can be arbitrarily chosen as long as the maximum number of consecutive non-sensing nodes also scales as $O(\sqrt{n})$.
\end{abstract}

\section{Introduction}
Consider a scenario where a source holds a time-varying information that is crucial to a user or a group of users. The users need to track this information as closely as possible in real-time to achieve their objectives. The age of information (AoI) \cite{kaul2011minimizing, yates21agesurvey, Kosta17agesurvey, Sun19agesurvey} is a novel metric introduced to quantify how fresh this information is within communication systems. Specifically, AoI measures the time that has elapsed since the last received update was generated at the source, allowing to evaluate how current the information is at the receiver's end. 

This metric has garnered significant attention as the demand for real-time data processing and timely information delivery have become critical in various applications, such as, sensor networks, autonomous systems, and the internet of things (IoT) in 6G systems. For instance, in the context of autonomous vehicles, to prevent collisions and enhance overall traffic safety, it is essential that multiple vehicles receive real-time data about the positions of other vehicles. AoI is particularly important in systems where the relevance of information diminishes over time, such as monitoring networks, where outdated data can lead to incorrect decisions or system inefficiencies. 

\begin{figure}[t]
    \begin{center}
    \subfigure[]{%
    \includegraphics[width=0.49\linewidth]{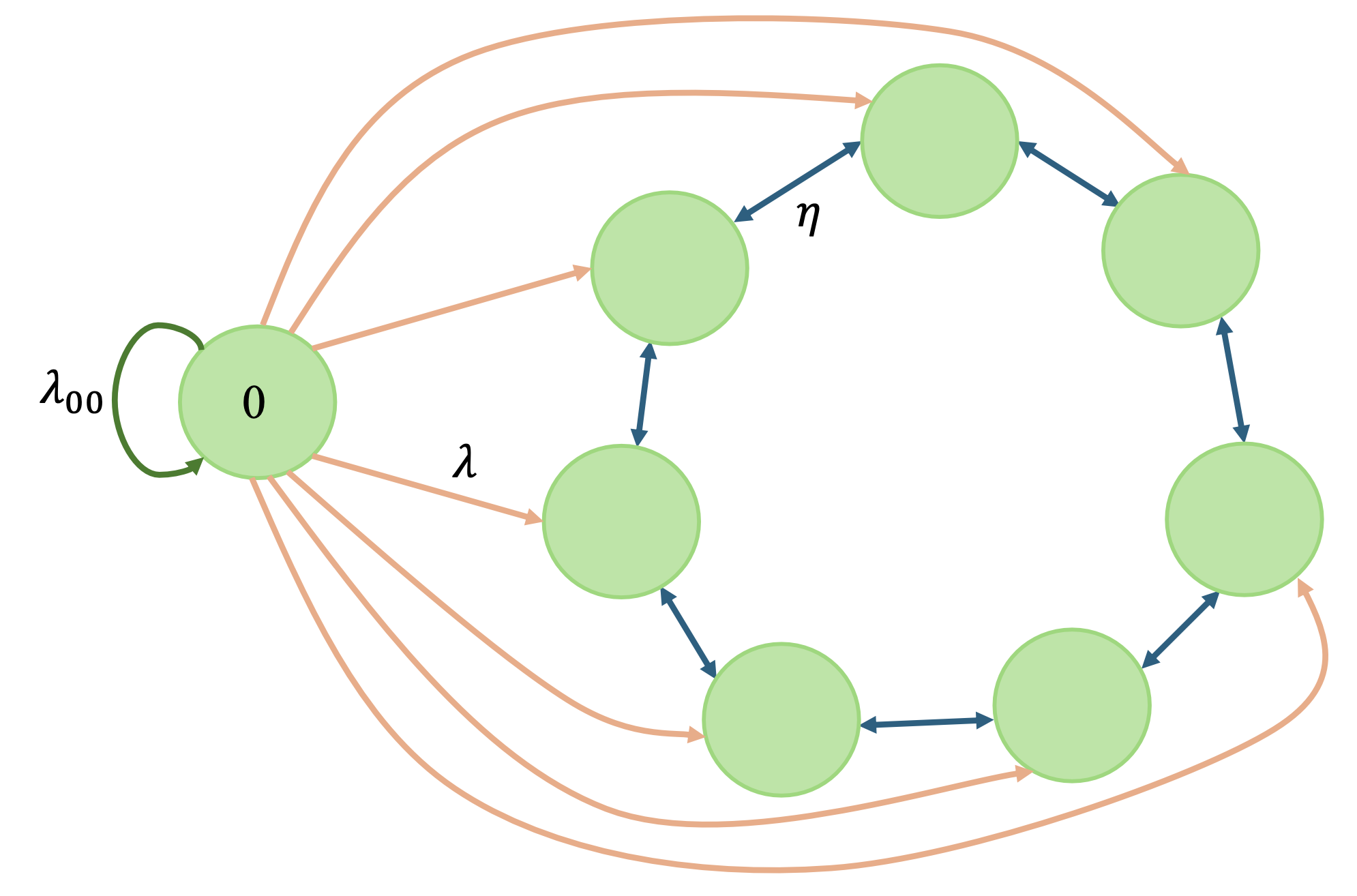}}
    \subfigure[]{%
    \includegraphics[width=0.49\linewidth]{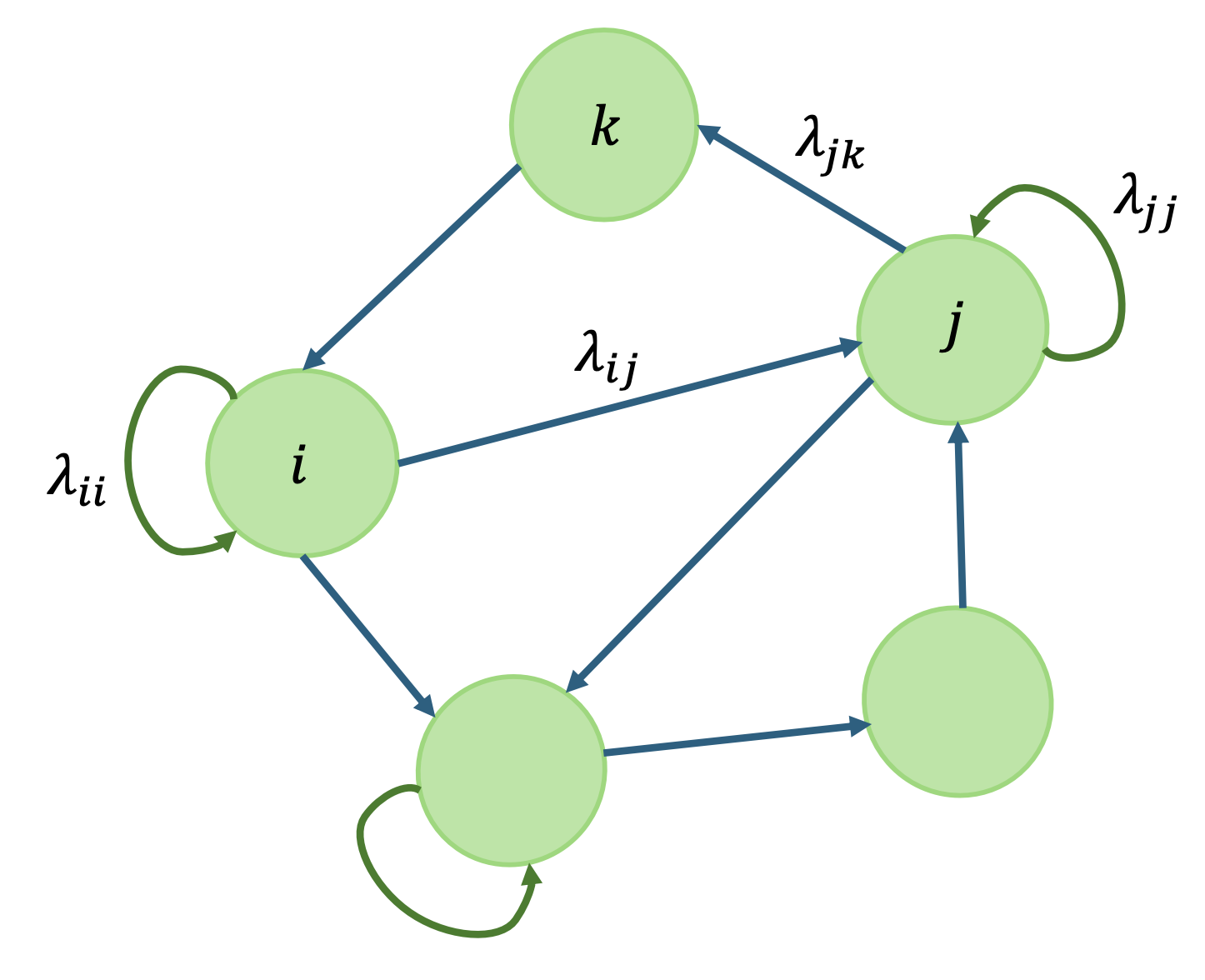}}
    \end{center}
    \vspace*{-0.2cm}
    \caption{(a) An example single source (single sensing node) network, and (b) an example multiple sensing node network.}
    \label{single_multi_source}
\end{figure}

Another significant advancement in this domain is the concept of \emph{version age}, where freshness of information is measured by a discrete quantity, i.e., version of the last update, instead of the absolute time \cite{yates2021versionage, Eryilmaz21, bastopcu20_google}. Every time a new reading from the source becomes available to a node (that is the ``source node'' in the current state-of-the-art), the version age of all nodes in the network is incremented with the exception of the source node. The performance of version AoI has been analyzed across different network architectures, such as fully connected networks \cite{yates2021versionage}, ring networks \cite{yates2021versionage, baturalp21comm_struc}, and grid networks \cite{srivastava2023age}; see \cite{aoi_gossip_litsurvy} for a recent survey. However, despite these advances, the current literature only deals with a single source node in the network (as shown in Fig.~\ref{single_multi_source}(a)) which generates all the updates according to a Poisson process.

In this paper, we build on a novel interpretation of the source node in the current literature, and introduce a new model. The so-called source node can be understood as a node of the network which in fact samples the information source, possibly with continuous distribution, such as the temperature or location, at times that follow a Poisson point process (PPP). Given this interpretation, one can immediately recognize that there is no need to limit such a node with sensing capability to be unique. Indeed, we consider having \emph{multiple views} of the information source via a subset of the nodes in the network (see Fig.~\ref{single_multi_source}(b)), which we call the \emph{sensing nodes}. 

For instance, consider a fleet of drones deployed to monitor a rapidly changing environment, such as during a disaster response operation. Each drone is responsible for gathering critical data, such as temperature, air quality, or the location of people in need of rescue, within the same area and relaying the most recent information back to a central user or system. In this paper, we will consider the scenario where there is a single underlying source, but there are multiple sensing nodes. Therefore, we will make a distinction between a source node in the prior art, where the underlying random source and the unique sensing node were both called the ``source node'', here we will call the underlying random source as the source and the nodes which sample/sense it the sensing nodes.  

In comparison between single- and multiple-view scenarios, we note that in a multiple-view scenario: (i) the AoI for a particular sensing node might be non-zero because another sensing node could have more up-to-date information, and (ii) the sensing node can also receive updates from other nodes of the network whether they are sensing or not-a-sensing nodes. We calculate the average version AoI for a network with multiple sensing nodes and demonstrate that the scaling behavior of the version AoI, in two well-known extreme cases of the system networks, i.e., fully connected networks and ring networks, is consistent with that observed in these networks with single-views under similar network configurations. 

The scenarios with multiple source updates have also been considered, e.g., see \cite{yates2018age, javani2021age}. However, analysis of the multiple-source scenarios is quite different from the treatment of the network with multiple views of a single source that we aim to analyze in this work. 

The rest of the paper is organized as follows. In Section~\ref{sec2:prelim}, we introduce the preliminaries, including definitions of version and version AoI for a multiple sensing nodes structure. In Section~\ref{sec3:main}, we present our proposed distributed sensing structure. Specifically, in Section~\ref{sec3_1:derivation}, we derive the average version AoI for the proposed scheme. Then, in Sections~\ref{sec3_2:ring}, \ref{sec3_3:line}, and \ref{sec3_4:fully}, we analyze the scaling behavior of the version AoI for three different network structures: ring networks, line networks, and fully connected networks, respectively.

\section{Preliminaries} \label{sec2:prelim}
We consider a network of $n$ nodes labelled from 1 to $n$, i.e., $[n] = \{1, 2, \ldots, n\}$. In this network, e.g., a sensor network, an accurate clock between the nodes may not be available, and consequently, the timeline at each node is measured by version updates. The set of all nodes is denoted by $\mathcal{N}$. A subset of the nodes $I \subset \mathcal{N}$ sense a time-varying source. To incorporate most physical phenomena such as location, temperature, air quality, and so on, we consider a source with a continuous distribution in general. A view of the source is available to node $i$,  $i \in I$,  by direct sampling of the source at times that is determined by a Poisson point process (PPP) of rate $\lambda_{ii}$ denoted in short hand notation as PPP$(\lambda_{ii})$. Any update from a node $i \in I$ constitutes a new version of the information that is available to the whole network. The information at node $i \in \mathcal{N}$ is shared with node $j$ in the network at times that are determined by a PPP$(\lambda_{ij})$. We note that the nodes in set $I$ may also receive an update of the source through the network from an adjacent node. By definition, the nodes and their updates through the PPP$(\lambda_{ij})$ for $i, j \in \mathcal{N}$ form a network that can be represented by a directed graph where the weight of the edge from node $i$ to node $j$ is given by $\lambda_{ij}$. For simplicity of derivation and presentation of the results, in this paper, we restrict our attention to the case that all PPPs are independent.   

Let us define the current version of the source sample available at node $i$ and time $t$ by $V_i(t)$. We note that $V_i(t)$ could merely be an abstract value that is used for calculation of the version AoI and it is not necessarily known to the nodes. However, one can consider a shared memory that is accessible to the nodes which contains the latest version of the source sample known to the whole network. In which case, when a sensing node performs a new sampling of the source, it increments this shared memory and marks its new packet with a time stamp and the new version. 

Due to the stochastic nature of the PPP used in sampling at the nodes in $I$ and communications between the nodes in the network, $V_i(t)$ is also a random process which possibly has discontinuities only at the points that one of the PPP$(\lambda_{ij})$, $i,j \in \mathcal{N}$ occurs, where at such a point, $V_i(t^-)$ and $V_i(t)$ indicate the value of $V_i$ before and after the possible update, respectively. We assume that the nodes use gossip to send their updates to their neighbors. The node $i$ sends the update to the node $j$ when the event corresponding to PPP$(\lambda_{ij})$ happens. The update from a node is composed of the current version of the information at that node and its corresponding time stamp. The time stamp is added to the information packet along with the source sample at each sensing node, i.e., the nodes $i \in I$. Each node only keeps the latest view of the source available at that node. Hence, the receiving node $j$ updates its version to the latest available version, i.e., $V_i(t^-)$ is replaced by $V_i(t) = \max (V_i(t^-), V_j(t))$. To clearly distinguish the value of $V_i(t)$ before and after update, we use the notation $V_i(t^-)$ to denote the available version to the node $i$ right before the update from node $j$ is applied. Let $\pi_{ij}(Y(t))$ denote the operator which provides the value of a random process $Y(t)$ at time $t$ after the PPP$(\lambda_{ij})$ has occurred. Using this notation, we write $V_i(t) = \pi_{ij}(V_i(t)) = \max (V_i(t^-), V_j(t))$. Without loss of generality, we can assume that the two PPPs do not happen at exactly the same time and differ by at least a small $\epsilon$ time difference due to the continuous (exponential) distribution between the points and the discrete nature of the PPP.

The version for a subset of the nodes $S \subset \mathcal{N}$ at time $t$ is defined as $V_S(t) = \max_{i \in S} V_i(t)$. Formally the version of the network at time $t$ is defined as $V_{\mathcal{N}}(t)$ which is the best (highest) version that can be available at any node in the network. We define the AoI at node $i$ as $X_i(t) = V_{\mathcal{N}}(t) - V_i(t)$. Hence, the version age of information (version AoI) at a subset $S \subset \mathcal{N}$ may be written as $X_S(t) = \min_{i \in S} X_i(t)$. 

\section{Main Result} \label{sec3:main}
In this section, we borrow the results from well-developed theory of stochastic hybrid and switched systems (SHS). This theory has applications mainly in modern control theory, where the control system comprises a hybrid of continuous and discrete time components, inputs (control actuators), and outputs. In this paper, we aim to characterize the average version AoI $v_S(t) = \E [X_S(t)]$, particularly, for an arbitrary node $i$, i.e., $S = \{i\}$. The application in our problem is immediate, as we are interested in average (in continuous time) version age of information (which takes discrete values corresponding to discrete version updates) and Poisson point process which happens in discrete time intervals with exponential distribution (continuous) between its sample points.

\subsection{Derivation of Average AoI} \label{sec3_1:derivation}
For a piece-wise linear SHS system where the derivative of the measurement function $Y(t)$ only has discontinuities as a result of a jump through a point process in time, the measurement function satisfies Dynkin's formula \cite{yates2012real},
\begin{align}\label{eq:derivative1}
    \frac{d }{dt} \E[Y(t)] = \E\bigg[ \sum_{(i,j) \in \L} \lambda_{ij} \big( \pi_{ij}(Y(t)) - Y(t^-) \big) \bigg],
\end{align}
where $Y(t^-)$ denotes the value of $Y$ right before the update at time $t$. For $Y(t) = X_S(t)$, we have, 
\begin{align}\label{eq:pi}
    \pi_{ij}(X_S(t)) = \begin{cases} X_S(t^-) + 1, & j=i, j \in I, j \not \in S, \\ 
    0, & j = i, j \in I, j \in S, \\ 
    X_S(t^-), & j = i, j \not \in I, \\ 
    X_S(t^-), & j \neq i, j \not \in S, \\ 
    X_S(t^-), & j \neq i, j \in S, i \in S, \\ 
    X_{S \cup \{i\}}(t^-), & j \neq i, j \in S, i \not \in S.
    \end{cases} 
\end{align}
Combining \eqref{eq:pi} for $Y(t) = X_S(t)$ with \eqref{eq:derivative1}, the right hand side of \eqref{eq:derivative1} can be rewritten as,
\begin{align}\label{eq:deri}
    \frac{d}{dt} \E[X_S] = & \E \bigg[ \sum_{j \not \in S, j \in I} \lambda_{jj} (X_S + 1 - X_S) - \sum_{j \in S, j \in I} \lambda_{jj} X_S \nonumber \\ 
    &   \quad + \sum_{j \in S, i \not \in S, j \neq i} \lambda_{ij} (X_{S \cup \{i\}} - X_S) \bigg].
\end{align}
We note that $\lambda_{jj} = 0$ where $j \not \in I$, hence, the corresponding term in calculation of \eqref{eq:derivative1} does not appear. Also, if $j \in S$ and $i \not \in S$ then trivially $j \neq i$, hence, the condition $j \in S, i \not \in S, j \neq i$ can be reduced to $j \in S, i \not \in S$. By setting $\frac{d}{dt} \E[X_S] = 0$ in \eqref{eq:deri}, the stationary point must satisfy
\begin{align}
     &\sum_{j \not \in S, j \in I} \lambda_{jj} - \sum_{j \in S, j \in I} \lambda_{jj} \E[X_S] \nonumber \\
     & \quad + \sum_{j \in S, i \not \in S, j \neq i} \lambda_{ij} (\E[X_{S \cup \{i\}}] - \E[X_S]) = 0,
\end{align}
and by setting $v_S = \E[X_S]$ and $v_{S \cup \{i\}} = \E[X_{S \cup \{i\}}]$ after rearranging, we have the following result.

\begin{proposition}\label{thm:thm1}
The average version AoI for a subset $S$ in a network with sensing nodes $i \in I$ that sample a view of the source at times characterized by PPP$(\lambda_{ii})$, and update received at node $j$ from node $i$ at the times given by PPP$(\lambda_{ij})$, $i \neq j$ is given by
\begin{align}\label{eq:main}
     v_S(t) = \frac{ \sum_{j \not \in S, j \in I} \lambda_{jj} + \sum_{j \in S, i \not \in S} \lambda_{ij} v_{S \cup \{i\}}}{\sum_{j \in S, j \in I} \lambda_{jj} + \sum_{j \in S, i \not \in S} \lambda_{ij}}.
\end{align}
\end{proposition}

Here, a few remarks are in order: First, we note that the average renewal rate for the whole network is given by $\sum_{j \in I} \lambda_{jj}$. In a network with a single sensing node $I = \{0\}$, the average renewal rate is $\lambda_{00}$. The formulation in \cite{yates2021versionage} derives the average AoI in a network with single source node $0$ with Poisson distribution $\lambda_{00}$ and $n$ nodes, $i \in [n]$. If we interpret the renewal of the view of the node $0$ as the arrival (or generation) of the new information by node $0$ directly as a source, the average AoI given by \eqref{eq:main} for the special case of single view of the source reduces to the result of  \cite[Thm.~1]{yates2021versionage}, in which case \eqref{eq:main} reduces to,
\begin{align}\label{eq:main-single}
     v_S(t) = \frac{ \lambda_{00} + \sum_{j \in S, i \not \in S} \lambda_{ij} v_{S \cup \{i\}}}{ \sum_{j \in S} \lambda_{0j} + \sum_{j \in S, i \not \in S} \lambda_{ij}}.
\end{align}

Second, the minimum average version AoI taken over all possible nodes of the network, $v_{\{i\}}$, $i \in \mathcal{N}$ achieves its minimum, that is equal to zero for $v_I = 0$, in the case of a network with single view at its unique viewing point $|I|=1$. However, the same is not true for the network with multiple views of the source. Example~\ref{ex:ex1} reveals a sharp contrast between a network with a single view of the source versus a network with multiple views of the source.

\begin{example} \label{ex:ex1}
    Consider the network in Fig.~\ref{fig:multi-view-ex}. 
    \begin{figure}[h]
        \centering
        \includegraphics[width=0.8\linewidth]{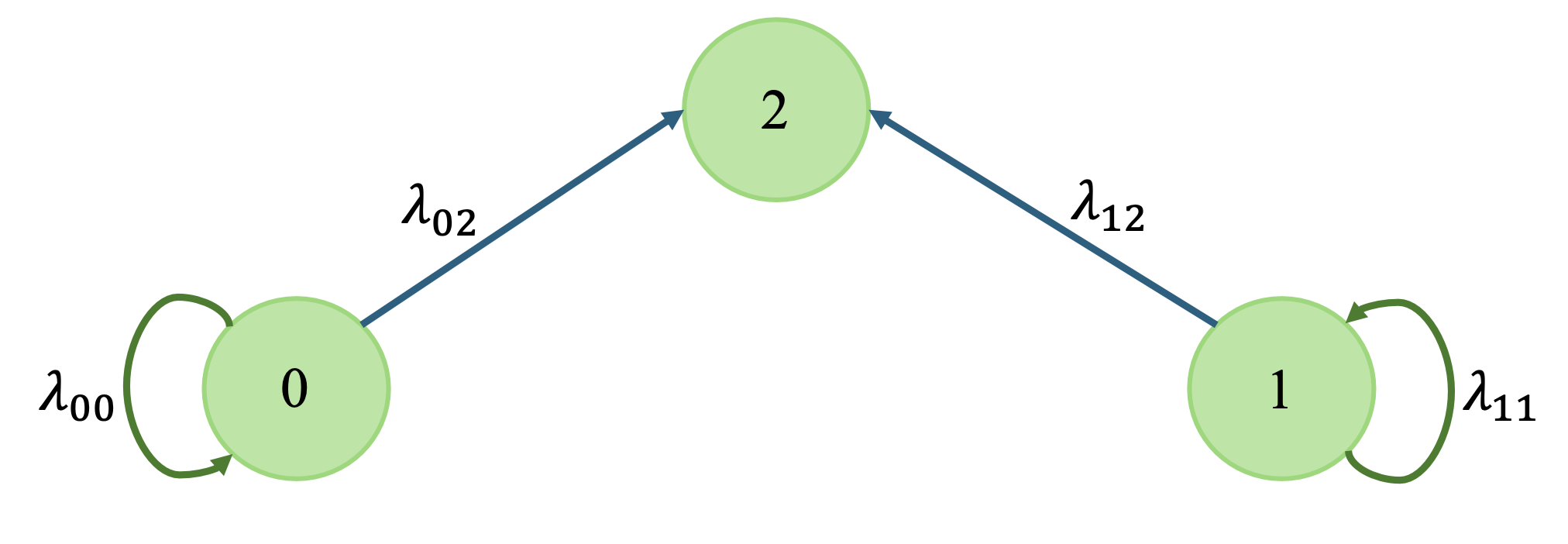}
        \caption{An example multi-view network with multiple sensing nodes. Nodes 0 and 1 are sensing nodes.}
        \label{fig:multi-view-ex}
    \end{figure}  
    
    Let us use the short hand notation $v_{ijk}$ by directly specifying the elements of the set $S = \{i, j, k\}$ in the subscript to denote $v_S$. We have $v_{012} = 0$, and
    \begin{align}
        v_2 = & \frac{\lambda_{00}+\lambda_{11}+\lambda_{12} v_{12}+\lambda_{02} v_{02}}{\lambda_{12}+\lambda_{02}}, \\
        v_{12} = & \frac{\lambda_{00}+\lambda_{02} v_{012}}{\lambda_{11}+\lambda_{02}}, \\
        v_{02} = & \frac{\lambda_{11}+\lambda_{12} v_{012}}{\lambda_{00}+\lambda_{12}}.
    \end{align}
    Hence, the average version AoI for each node is given by
    \begin{align}
        v_2 = & \frac{\lambda_{00}+\lambda_{11}}{\lambda_{12}+\lambda_{02}}+\frac{\lambda_{00}\lambda_{12}}{(\lambda_{12}+\lambda_{02})(\lambda_{11}+\lambda_{02})} \nonumber \\ 
        &+ \frac{\lambda_{11}\lambda_{12}}{(\lambda_{12}+\lambda_{02})(\lambda_{00}+\lambda_{12})}, \\
        v_{1} = & \frac{\lambda_{00}}{\lambda_{11}}, \\
        v_{0} = & \frac{\lambda_{11}}{\lambda_{00}}.
    \end{align}
\end{example}

If we choose $\lambda_{11} = \lambda_{00} = 1$ and  $\lambda_{12} = \lambda_{02} = 10$, in Example~\ref{ex:ex1}, we obtain $v_0 = v_1 = 1$, and $v_2 = 0.19$. Thus, Example~\ref{ex:ex1} shows that the average AoI in neither of the viewing points (sensing nodes) is zero. It also shows that the node with the minimum AoI is not necessarily one of the viewing points. We next state the following proposition.

\begin{proposition}\label{thm:thm2}
    In a network with multiple views, i.e., $|I| > 1$ and PPP of finite rate, the average AoI for any node is always nonzero.
\end{proposition}

Even though Proposition~\ref{thm:thm2} could provide a pessimistic observation of the network with multiple views, we note that the average version AoI can indeed be a constant, i.e., $O(1)$ even as the network size grows large. 

The average AoI in a network provides a way to compare the freshness of information at different parts of the network. In a network with a single view of the source, the average AoI of the node indeed provides additional insight when the transition probabilities between the nodes change, i.e., when the network topology evolves. However, if the sampling rate for the view of the source changes from an instance of the network to another, the average AoI of a nodes does not directly provide a fair comparison to the freshness of the information at this node. Moreover, in comparing two networks with different numbers of viewing points, the average AoI is a fair measure if the total renewal rate of the network in comparison is a constant. Hence, we propose to use the \underline{No}rmalize a\underline{V}erage \underline{A}o\underline{I}, i.e., NoVAI, defined as the ratio of the average AoI to the average renewal rate for the view of the source by all nodes in the network as a fair measure. NoVAI can also provide a fair measure in comparison between two networks, when the number of sensing nodes changes in the network, or when the sampling rates of the sensing nodes change. 

In the following, we investigate the scaling of the average version AoI in the network for various important network typologies, namely, ring, line, and fully connected networks. We present the results for the average version AoI of any single node in the network, as in \cite{yates2021versionage}. Other measures such as the average version AoI taken over all single nodes of the network has also been considered \cite{kaswan2022age} which does not necessarily enforce the scaling on any single node.

\subsection{Ring Network} \label{sec3_2:ring}
\begin{figure}[t]
    \begin{center}
    \subfigure[]{%
    \includegraphics[width=0.49\linewidth]{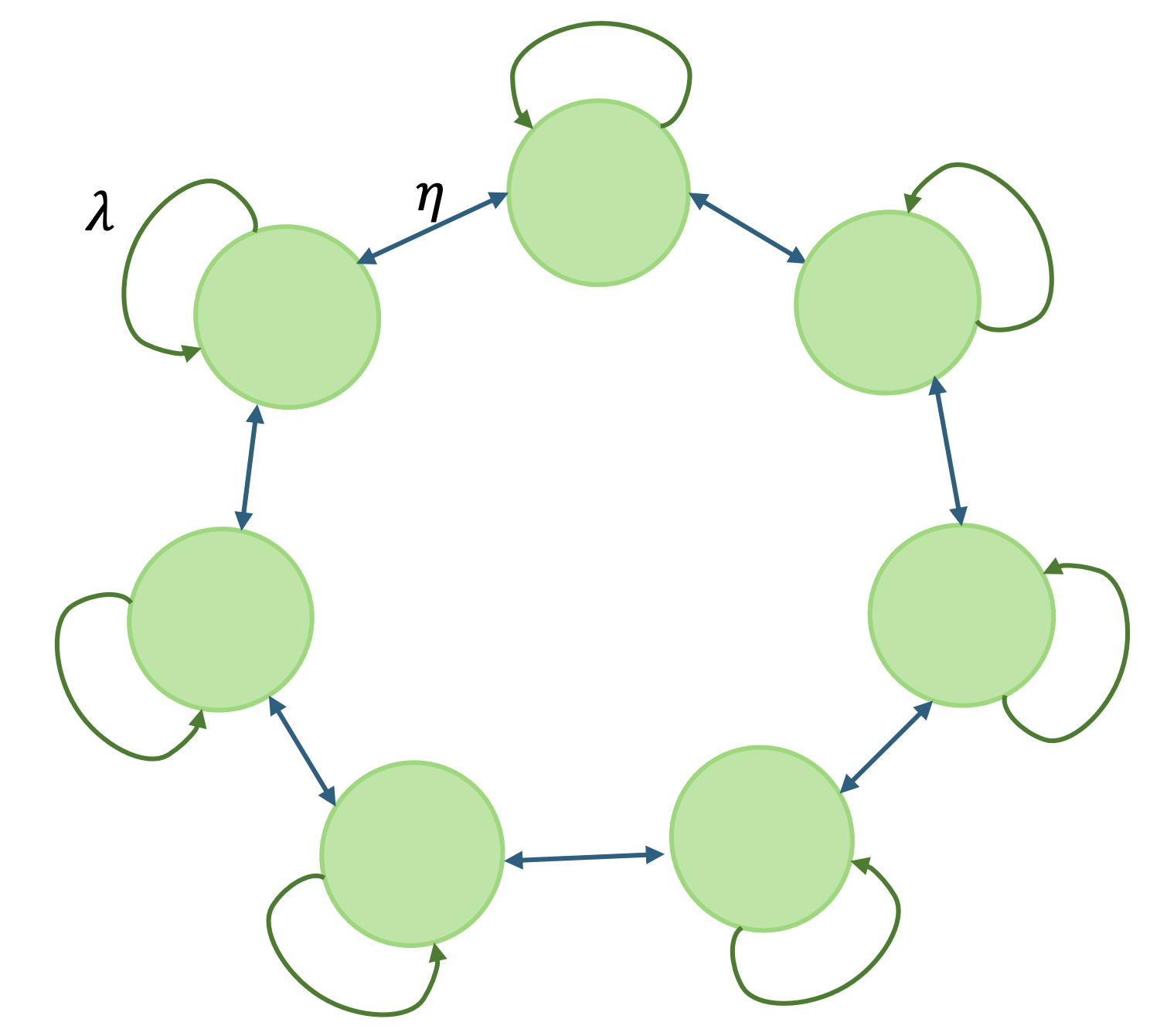}}
    \subfigure[]{%
    \includegraphics[width=0.49\linewidth]{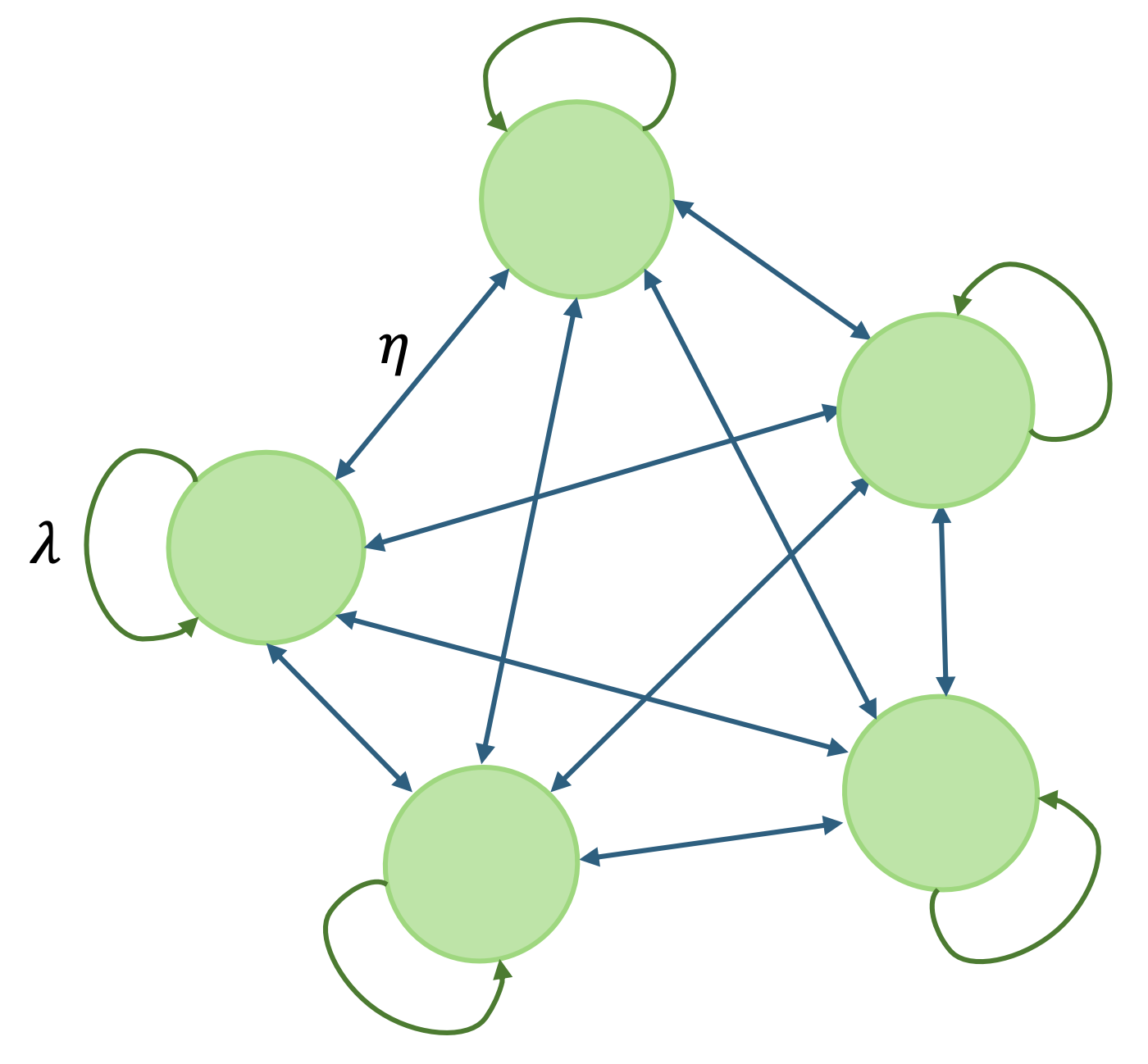}}
    \end{center}
    \vspace*{-0.2cm}
    \caption{(a) Ring network, and (b) fully connected network, where all nodes are sensing nodes.}
    \label{fig:ring_fully_nework}
\end{figure}

Consider a network of $n$ nodes labeled as $i \in [n]$. Using PPP$(\lambda)$, the node $i$ receives a renewal sample of the information source and keeps the last version and its time stamp; see Fig.~\ref{fig:ring_fully_nework}(a). The node $i$ also receives updates from adjacent nodes $(i-1) \mod n$ and $(i+1) \mod n$ at a rate governed by PPP$(\eta)$ and keeps the version with the latest time stamp. Let $v_k$ denote the average AoI for a connected subset of nodes of size $k$, i.e., the nodes with indices $j+1 \mod n, j+2 \mod n, \ldots, j+k \mod n$. Due to the symmetry of the network, $v_k$, $k\in [n]$ is well defined as any subset of $k$ connected nodes are equal. Using \eqref{eq:main} we can write $v_k$ in terms of $v_{k+1}$ as
\begin{align}
    v_k = \frac{(n-k) \lambda + 2 \eta v_{k+1}}{k \lambda + 2 \eta}.
\end{align}
Defining $\beta = 2 \eta / \lambda$, we have
\begin{align}
    v_k = \frac{(n-k)}{k+\beta} + \frac{\beta}{k+\beta}   v_{k+1}.
\end{align}
Using this recursion, we find the AoI for a single node $v_1$ as
\begin{align}
    v_1 &= \frac{(n-1)}{1+\beta} + \frac{\beta}{1+\beta} \bigg( \frac{(n-2)}{2+\beta} + \frac{\beta}{2+\beta} \bigg( \cdots \\
        &=  \Gamma(1+\beta) \sum_{k=1}^{n-1} \frac{n-k}{\Gamma(1+k+\beta)}.
\end{align}
For $\lambda = \lambda_{00}/n$, $\eta = \eta_{00}/2$, and $\lambda_{00} = \eta_{00}$, we have $\beta = n$, and using the fact that $v_n = 0$, we have
\begin{align}
    v_1 &=  \sum_{k=1}^{n-1} \frac{n^{k-1}(n-k)}{\prod_{i=1}^k(n+i)} = n! \sum_{k=1}^{n-1} \frac{n^{k-1}(n-k)}{(n+k)!} \\
        &\leq  \frac{n!}{n^n} \sum_{k=1}^{n-1} \frac{n^{n+k}}{(n+k)!} \leq\frac{n!}{n^n} e^n.
\end{align}
Using Sterling's formula for large $n$, $n! \sim \sqrt{2 \pi n} \frac{n^n}{e^n}$, we conclude that $v_1 \sim O(\sqrt{n})$. Even though we derived this scaling results for the network with distributed views, following similar derivation, we can derive the same scaling results for the network with an additional node 0 connected to the $n$ nodes on the ring with the link following the activation through PPP$(\lambda/n)$. This scaling of $O(\sqrt{n})$ was conjectured in \cite{yates2021versionage} as experimentally observed in \cite[Fig.~4]{yates2021versionage}, and mathematically proved to be true in \cite[Lemma~2]{baturalp21comm_struc} for the single view case. The above derivation proves it for the multi-view case.  
 
\subsection{Line Network} \label{sec3_3:line}
The line network is composed of $n$ nodes labeled as $i \in [n]$ where the node $1 < i < n$ only receives updates from adjacent nodes $(i-1)$ and $(i+1)$ and node 1 and $n$ only receive updates from node 2 and $n-1$, respectively, at a rate governed by PPP$(\eta)$ and keeps the version with the latest time stamp. Consider a general case where there are only $q$ sensing nodes which sample the source corresponding to PPP$(\lambda)$. Let $d$ be the maximum number of consecutive nodes where none of the $i$ is a sensing node, i.e., we have $i \not \in I$ for any node $i$ in this set of consecutive nodes.

Let $v_{j,k}$ denote the average version AoI for a connected subset of nodes of size $k$ which starts with the node index $j+1$, i.e., the nodes with indices from the set $\mathcal{S}_{j,k} = \{(j+1) \mod n, (j+2) \mod n, \ldots, (j+k) \mod n \}$. Since the position of the sampling points is not known, the value of $v_k$ is indeed a function of $j$, however, we can define an upper bound on $v_{j,k}$ for all $k$ as $\overline{v}_k$ which is obviously not less than $\max_j v_{j,k}$, i.e., maximum of average AoI over all possible connected subset of nodes. In the following, we derive recursive expression for $\overline{v}_k$. We use the following properties to reduce the number of recursive expression for $\overline{v}_k$. (i) We note that by starting from the largest set, i.e., $k=n$, we have $\overline{v}_n = \overline{v}_\mathcal{N} = 0$. (ii) For any $k<n$, we note that $\overline{v}_k$ only depends on $v_{j,k+1}$ and if in any such expression $v_{j,k+1}$ is replaced by $\overline{v}_{k+1}$ then $\overline{v}_k$ is not reduced and still is an upper bound for $v_{j,k}$ for all $j$. (iii) The worst case update for any of such expressions for update from the $v_{j,k}$ is also when it comes from a single node with PPP$(\eta)$. This statement is intuitively true, and the formal proof relies on the fact that for any $T \subset S \subset \mathcal{N}$, we have $v_T \geq v_S$, and therefore $\overline{v}_{k+1} < \overline{v}_{k}$. (iv) Consider $v_{j,k}$ for any $j,k$. There is $n-k$ nodes outside $\mathcal{S}_{j,k}$ out of which at least $\lfloor \frac{n-k-q}{d+1} \rfloor^+$ nodes are sensing nodes, where $\lfloor x \rfloor^+$ is non-negative part of the floor $x \in \mathbb{R}$. For example, when $k = n-2q$, there could be exactly $q$ non-sensing nodes in each side of the set $\mathcal{S}_{j,k}$, but if $k = n - 2q -1$ there is at least one sensing node in the set $\mathcal{S}_{j,k}$. Hence,  using \eqref{eq:main} and the above reduction technique, we have 
\begin{align} 
    \overline{v}_k = & \frac{(q-\lfloor \frac{n-k-q}{d+1} \rfloor^+) \lambda + \eta \overline{v}_{k+1}}{\lfloor \frac{n-k-q}{d+1} \rfloor^+ \lambda + \eta}, \quad k \geq n-q-md, \label{eq:recurssive_line1}\\
    \overline{v}_k = & \frac{k \lambda + \eta \overline{v}_{k+1}}{(q - k) \lambda + \eta}, \qquad \qquad \qquad \quad k < n-q-md, \label{eq:recurssive_line2}
\end{align}
where $n-q = dm + p$, $p < d$ for some positive integers $m, p$. Using definition of $d$ and $q$ it is immediate that $(d+1)q \geq n$, otherwise it contradicts the fact that $d$ is the maximum number of (not sensing) nodes between any two consecutive sensing nodes. The worst case happens when $(d+1)q \geq n$ which means that $m=q$ and $p=0$. Define $\beta = \eta / \lambda$, the expression for $\overline{v}_1$ is given by
\begin{align}
    \overline{v}_1 = \frac{2qd}{\beta} + \sum_{j = 1}^{m-1} \sum_{i = 1}^{d+1} \frac{(q-j) \beta^{(j-1)(d+1)+i-1}}{(1+\beta)^i \left( \prod_{l=1}^{j-1} (l+\beta) \right)^{d+1} }.
\end{align}
Using geometric series sum, we have
\begin{align}
    \overline{v}_1 &\leq \frac{2qd}{\beta} + q + \frac{q}{\beta} \sum_{j = 1}^{m-1} \left( \frac{\beta^j}{\prod_{l=1}^{j} (l+\beta)} \right)^{d+1} \\
    &= \frac{2qd}{\beta} + q + \frac{q\Gamma(1+\beta)^{(d+1)}}{\beta^{\beta(d+1)+1}} \sum_{j = 1}^{m-1} \left( \frac{\beta^{(j + \beta)}}{\Gamma(1+j+\beta)} \right)^{d+1}. 
\end{align}
Using gamma function approximation $\Gamma(x) = x^x \sqrt(2 \pi x)/ e^x$, and using the fact that $\lambda$ should be scaled down to $\lambda/q$, we find that the upper bound scales as $\overline{v}_1 \sim O(d+q)$. It is immediate that by converting line network to a ring by adding a link between the two ends of the line, the average AoI for any subset would not be worse, hence the scaling results in the line network will hold true in the ring network.

The asymptotic behavior predicted for the upper bound reveals that in order to achieve $O(\sqrt{n})$ bound on the average AoI of any single node in the line network, it is enough to have $q \sim O(\sqrt{n})$ and $d \sim O(\sqrt{n})$. We note that the asymptotical bound is not necessarily tight in all cases and may even fall short of some simple bounds. Nonetheless, the achievement of $O(\sqrt{n})$ for average AoI of any single node in the network with only $q \sim O(\sqrt{n})$ at any position is quite interesting, while the best known result for a ring network requires $q \sim O(n)$ \cite{yates2021versionage}. We note that scaling of average AoI is also a function of scaling of $d$ and the condition $d \sim O(\sqrt{n})$ is crucial to achieve $O(\sqrt{n})$ scaling for the average version AoI. 

Two notes are in order here: First, the bound is not a function of particular structure or pattern for the placement of the sensing nodes. Second, this bound may be used to emphasize the importance of the distributed or multi-view of a source. More specifically, if a single node, say in the middle of the line network or any node on a ring, samples the source corresponding to PPP$(\lambda_0)$, the average AoI per node scales as $O(n)$, while by using distributed sensing with $q = \lceil \sqrt{n} \rceil$ that sample the source corresponding to PPP$(\lambda_0/q)$ collectively at the same rate $\lambda_0$, the average AoI per node scales as $O(\sqrt{n})$. Comparison of the average AoI scaling per node for various scaling of $d$ is shown in Fig.~\ref{fig:effect_d_AoI} and discussed in the following.

\begin{figure}[t]
    \centering
    \includegraphics[width=0.9\linewidth]{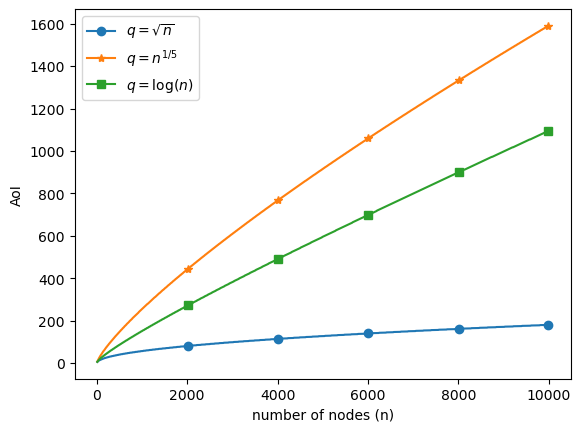}
    \caption{Effect of changing $q$ in average AoI in line/ring network.}
    \label{fig:effect_q_AoI}
    \vspace*{-0.3cm}
\end{figure}

\begin{figure}[t]
    \centering
    \includegraphics[width=0.9\linewidth]{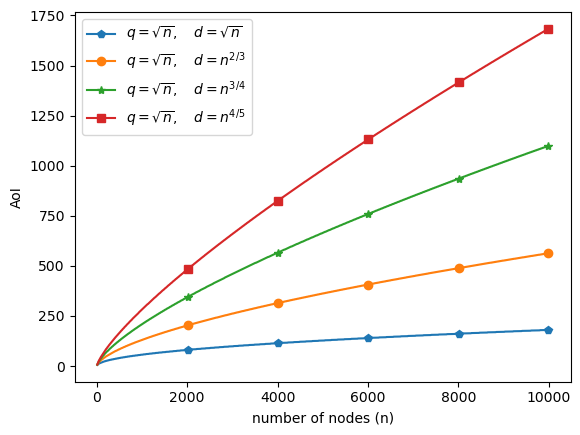}
    \caption{\label{fig:effect_d_AoI} Effect of changing $d$ in average AoI in line/ring network.}
    \vspace*{-0.3cm}
\end{figure}

Fig.~\ref{fig:effect_q_AoI} illustrates the numerical evaluation of the upper bound $\overline{v}_1$ in \eqref{eq:recurssive_line1}-\eqref{eq:recurssive_line2} for different scaling of the $q$ and $d$. We observe that if the number of sensing nodes $q$ scales as $O(\sqrt{n})$, $O(\log(n))$, and $O(n^{1/5})$ and $d$ accordingly scales as $O(n/q)$, i.e., $O(\sqrt{n})$, $O(n / \log(n))$, and $O(n^{4/5})$, the corresponding curve scales as $O(d+q)$, i.e., $O(\sqrt{n})$, $O(n / \log(n))$, and $O(n^{4/5})$, respectively. Moreover, we note that $q(d+1) \geq n$ forces $d \geq -1 + n/q$ and on the other hand we have $n\geq q+d$ which provides an upper bound on $q$ as $d \leq n-q$ which means that $O(qd)$ can be indeed larger than $O(n)$. Fig.~\ref{fig:effect_d_AoI} illustrates the bound \eqref{eq:recurssive_line1}-\eqref{eq:recurssive_line2} for different values of $d$ and $q$ and the numerical values show that the scaling for large $n$ is $O(d+q)$. In particular, even though $q \sim O(\sqrt{n})$, as $d$ scales with $O(n^{2/3})$,$O(n^{3/4})$, and $O(n^{4/5})$, the upper bound on the average AoI per a single node, i.e., $\overline{v}_1$ scales as $O(n^{2/3})$, $O(n^{3/4})$, and $O(n^{4/5})$, respectively.

\subsection{Fully Connected Network} \label{sec3_4:fully}
Consider a fully connected network of $n$ nodes labeled as $i \in [n]$. As shown in Fig.~\ref{fig:ring_fully_nework}(b), each node samples the information source using PPP$(\lambda)$ and only keeps the last view of sample from the information source as the current version with its time stamp. Each node receives the version available at another node governed by PPP$(\eta)$ and only keeps the version with the latest time stamp. We note that in general $\eta$ and $\lambda$ could indeed be a function of $n$, e.g., $\eta = \eta_0/n$, $\lambda = \lambda_0/n$ and so on. Let $v_k$ denote the average version AoI for a subset of nodes of size $k$. Due to the symmetry, $v_k$, $k\in [n]$ is well defined as any subset of $k$ nodes are equal. Using \eqref{eq:main} we write $v_k$ in terms of $v_{k+1}$ as
\begin{align}
    v_k = \frac{(n-k) \lambda + (n-k) k \eta v_{k+1}}{k \lambda + (n-k) k \eta}.
\end{align}
Define $\alpha = \lambda / \eta$, we have
\begin{align}
    v_k = \frac{(n-k)}{n-k+\alpha} \bigg( \frac{\alpha}{k} +  v_{k+1}\bigg).
\end{align}
Using the above recursion, we can find the AoI for a single node $v_1$ as
\begin{align}
    v_1 = & \frac{(n-1)}{n-1+\alpha} \bigg( \frac{\alpha}{1} +  \frac{(n-2)}{n-2+\alpha} \bigg( \frac{\alpha}{2} +  \cdots \nonumber \\
    & + \frac{1}{1+\alpha} \bigg( \frac{\alpha}{n-1} +  v_{n}\bigg) \cdots \bigg)
\end{align}
We note that $v_n = 0$, hence for $\alpha = 1$, i.e., $\lambda = \eta$ we have
\begin{align}
    v_1 &= \frac{(n-1)}{n} \bigg( 1 +  \frac{(n-2)}{n-1} \bigg( \frac{1}{2} +  \cdots + \frac{1}{2} \bigg( \frac{1}{n-1}\bigg) \bigg)  \\
    &= \sum_{k=1}^{n-1} \frac{n-k}{nk} \leq \log(n)
\end{align}

This means that if both $\lambda$ and $\eta$ scale as $O(\frac{1}{n})$, e.g., $\lambda = \lambda_{00}/n$ and $\eta = \eta_{00}/n$, the AoI for a single node in the network scale as $O(\log(n))$. Note that the same scaling is true if, e.g., $\eta = \frac{\eta_{00}}{n-1}$. This means that for fully connected network, the scaling of AoI using distributed sampling at the nodes is the same as that of the case where the version updates from a single view node are provided to the nodes of the network at the similar rate as derived in \cite{yates2021versionage}. In other words, there is no loss in terms of AoI if we perform distributed sampling. Note that the average renewal of the information in the network in both cases is equal to $\lambda_{00}$.

However, we note that if for some constants $\lambda_{00}$ and $\eta_{00}$, if $\lambda = \lambda_{00}/n$ and $\eta = \eta_{00}$ then the average AoI scales as $O(\frac{\log(n)}{n})$. Also, if $\lambda = \lambda_{00}/n$ and $\eta = \eta_{00}/n^2$, the average AoI scales as $O(n\log(n))$. These results are true for either of case of a network with multiple view or single view as in \cite{yates2021versionage}. This emphasizes the fact that the scaling of AoI varies as the scaling behavior of the mean of PPP for the links of the network varies. Hence, even though not stated directly in the theorem statement, the strong result of \cite[Thm.~1]{yates2021versionage} should be carefully interpreted only for the case that the mean of the PPP for the link of the network is of order $o(1/n)$. 

\section{Conclusion}
We introduced distributed sensing for version AoI in networks. We derived iterative expressions for the average AoI for any subset of the nodes. We established the scaling of the average AoI as the number of nodes in the network grows for multiple fundamental network topologies. 

\bibliographystyle{unsrt}
\bibliography{reference}

\end{document}